\documentclass[showpacs,preprintnumbers,amsmath,amssymb,superscriptaddress,floatfix,nofootinbib]{revtex4}
\usepackage{appendix}
\usepackage{graphicx}
\usepackage{epsfig}
\usepackage{epstopdf}
\usepackage{hyperref}
\usepackage{amsmath}
\usepackage{amsfonts}
\usepackage{amssymb}
\usepackage{mathtools}
\usepackage{multirow}
\usepackage{upgreek}
\usepackage{booktabs} 
\usepackage{subfigure}
\usepackage[justification=raggedright]{caption}
\usepackage{bm}
\usepackage{color}
\graphicspath{{./graph/}}
\usepackage{xcolor}
\usepackage{ulem}
\usepackage{dcolumn}

\newcommand{\be}{\begin{equation}}
\newcommand{\ee}{\end{equation}}
\newcommand{\ba}{\begin{eqnarray}}
\newcommand{\ea}{\end{eqnarray}}

\begin{document}
\renewcommand{\arraystretch}{1.5}

\title{Lattice QCD study of $\Lambda_c \Lambda_c$ scattering}
\author{\includegraphics[scale=0.2]{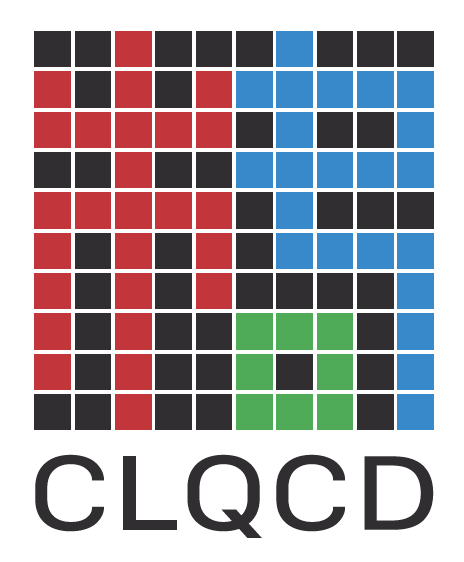}\\Hanyang Xing}
\affiliation{Institute of Modern Physics, Chinese Academy of Sciences, Lanzhou, 730000, China}
\affiliation{University of Chinese Academy of Sciences, Beijing 100049, China}

\author{Yiqi Geng}
\email{yqgeng@njnu.edu.cn}
\affiliation{Institute of Modern Physics, Chinese Academy of Sciences, Lanzhou, 730000, China}
\affiliation{Department of Physics and Institute of Theoretical Physics, Nanjing Normal University, Nanjing, Jiangsu 210023, China}

\author{Chuan Liu}
\affiliation{School of Physics, Peking University, Beijing 100871, China}
\affiliation{Center for High Energy Physics, Peking University, Beijing 100871, China}
\affiliation{Collaborative Innovation Center of Quantum Matter, Beijing 100871, China}

\author{Liuming Liu}
\email{liuming@impcas.ac.cn}
\affiliation{Institute of Modern Physics, Chinese Academy of Sciences, Lanzhou, 730000, China}
\affiliation{University of Chinese Academy of Sciences, Beijing 100049, China}

\author{Peng Sun}
\email{pengsun@impcas.ac.cn}
\affiliation{Institute of Modern Physics, Chinese Academy of Sciences, Lanzhou, 730000, China}
\affiliation{University of Chinese Academy of Sciences, Beijing 100049, China}

\author{Jiajun Wu}
\affiliation{University of Chinese Academy of Sciences, Beijing 100049, China}

\author{Zhicheng Yan}
\affiliation{Institute of Modern Physics, Chinese Academy of Sciences, Lanzhou, 730000, China}
\affiliation{Department of Physics and Institute of Theoretical Physics, Nanjing Normal University, Nanjing, Jiangsu 210023, China}

\author{Ruilin Zhu}
\affiliation{Department of Physics and Institute of Theoretical Physics, Nanjing Normal University, Nanjing, Jiangsu 210023, China}

\begin{abstract}
    We present the first lattice result of the near threshold $\Lambda_c\Lambda_c$ scattering with $I(J^P) = 0(0^+)$. The calculation is performed on two $N_f = 2+1$ Wilson-Clover ensembles with pion mass $m_\pi \sim 303$\,MeV and lattice spacing $a = 0.07746$\,fm. The L\"uscher's finite volume method is utilized to extract the scattering parameters from the finite-volume spectrum. The coupled channel $\Xi_{cc}N$ is ignored in the scattering analysis based on the observation that the energy levels computed from the $\Lambda_c\Lambda_c$ and $\Xi_{cc}N$ operators do not mix. The $\Sigma_c\Sigma_c$ channel is not included either since the energy range explored in this study is well below its threshold. Our results indicate that the interaction in the $\Lambda_c\Lambda_c$ single channel is repulsive, and the scattering length is determined to be $a_0 = -0.21(4)(8)$\,fm, where the first error is the statistical error and the second is the systematic error. 
  
\end{abstract}

\maketitle

\section{Introduction}
\label{sec:introduction}
Study of hadron-hadron interactions is an essential part of understanding the strong interactions and its underlying theory --- quantum chromodynamics(QCD). Among them, dibaryon systems contain rich dynamics in nature. The only known dibaryon bound state is the deuteron. A possible $SU(3)$ flavor-singlet bound state, the $H$-dibaryon composed of two $\Lambda$ baryons, was proposed long time ago \cite{Jaffe:1976yi}, but has not been observed in experiments yet. Comparing to the meson-meson scattering, lattice QCD study of baryon-baryon scattering is more challenging mainly due to the poor signal and the complexity in the contractions of dibaryon correlation functions. The lattice results on nucleon-nucleon scattering and the binding nature of the deuteron are still controversial as of today\citep{NPLQCD:2012mex, NPLQCD:2013bqy,  Berkowitz:2015eaa, Orginos:2015aya, Wagman:2017tmp, Horz:2020zvv, Amarasinghe:2021lqa}. Concerning the $H$-dibaryon, most of the lattice studies found attractive interactions between two $\Lambda$ baryons, but consensus on whether they can form a bound state and the magnitude of the binding energy remains elusive~\cite{Green:2021qol,Francis:2018qch,HALQCD:2019wsz,Yamaguchi:2016kxa,Beane:2011zpa,Shanahan:2011su}. Bound states of two heavy baryons are also predicted in some theoretical studies~\cite{Dong:2021bvy,Chen:2017vai,Lee:2011rka,Meguro:2011nr, Huang:2013rla,Oka:2013xxa,Garcilazo:2020acl,Li:2012bt,Gerasyuta:2011zx,Lu:2017dvm}. However, experimental data on two heavy baryon interactions is scarce since producing two heavy baryons in experiments is difficult. On the lattice side, there are a few studies on the scattering of two heavy baryons ~\cite{Lyu:2021qsh,Mathur:2022ovu}, where bound states of $\Omega_{ccc}-\Omega_{ccc}$ and $\Omega_{bbb}-\Omega_{bbb}$ are predicted, respectively. The ground state energy spectra of various heavy dibaryons are also investigated in lattice QCD~\cite{Junnarkar:2022yak,Junnarkar:2024kwd,Junnarkar:2019equ}.
In this study, we focus on the scattering of two $\Lambda_c$ baryons, analogous to the $H$-dibaryon but with the strange quarks replaced by charm quarks. 

The $\Lambda_c\Lambda_c$ scattering has been investigated in many theoretical studies, but the results are inconclusive. In studies such as Refs.\cite{Chen:2017vai, Gerasyuta:2011zx, Lu:2017dvm}, a bound state is identified in the single channel $\Lambda_c\Lambda_c$ scattering. Conversely, the results in Refs.\cite{Lee:2011rka, Dong:2021bvy, Carames:2015sya, Garcilazo:2020acl} do not support the existence of such a bound state. Other works suggest that $\Lambda_c\Lambda_c$ cannot form a bound state in a single channel, but coupling to $\Sigma_c \Sigma_c$ may lead to the formation of a bound state below the $\Lambda_c\Lambda_c$ threshold ~\cite{Meguro:2011nr, Huang:2013rla, Oka:2013xxa, Li:2012bt}. As a first-principle method, lattice QCD calculation of $\Lambda_c \Lambda_c$ scattering may provide crucial information for other theoretical studies. 

This work presents the lattice QCD calculation of $\Lambda_c\Lambda_c$ scattering based on the 2+1 flavor gauge ensembles with pion mass $m_\pi \sim 303$\,MeV and lattice spacing $a = 0.07746$\,fm. The L\"uscher's finite volume method~\cite{Luscher:1985dn,Luscher:1986pf,Luscher:1990ux} is employed to extract scattering information from the finite-volume spectrum. The scattering length and effective range of $\Lambda_c\Lambda_c$ scattering is obtained, and the results indicate repulsive interaction. The coupling with $N \Xi_{cc}$ and $\Sigma_c \Sigma_c$ is also discussed.

This paper is organized as follows. The details of the gauge ensembles and computational methods are introduced in Sec.~\ref{sec:ComputationalDetails}. In Sec.~\ref{sec:Spectrum}, we present the single particle spectrum and two-particle finite-volume spectrum. The scattering analysis and the results are given in Sec.~\ref{sec:Results}, followed by a summary in Sec.~\ref{sec:Summary}.

\section{Computational details}
\label{sec:ComputationalDetails}
The results presented in this paper are based on the gauge configurations generated by the CLQCD collaboration with 2+1 dynamical quark flavors using the tadpole improved tree level Symanzik gauge action and Clover fermion action ~\cite{CLQCD:2023sdb}. Numerous studies have been performed on these configurations, see e.g. ~\cite{Yan:2024yuq, Du:2024wtr, Liu:2023feb, Xing:2022ijm, Liu:2022gxf, Zhang:2021oja, Meng:2024gpd, Yan:2024gwp}. In this work we use two ensembles with the same pion mass $m_\pi \sim 303$\,MeV and lattice spacing $a = 0.07746$\,fm but with different volumes. The parameters of the two ensembles are listed in Table~\ref{tab:ensemble}. The valence charm quark mass is tuned to reproduce the physical spin-averaged mass of $\eta_c$ and $J/\Psi$, i.e. $\frac{1}{4}M_{\eta_c} + \frac{3}{4}M_{J/\Psi}$. The value of $m_{\pi} L$ is 3.81 and 5.72 for the two ensembles respectively. We expect that the volume corrections to the hadron energies, which are typically suppressed as $\sim e^{- m_{\pi}L}$, should not have noticeable impact on our final results. 
\begin{table}[htbp]
\begin{ruledtabular}
\centering
\begin{tabular}{cccccccccc}
    Ensemble   & $\beta$  & $a$(fm)   & $(L/a)^3\times T/a$  &$am_l$ &$am_s$  & $m_\pi$(MeV)   & $m_K$(MeV)     &$N_{conf}$\\
    \hline
    F32P30   & 6.41   &0.07746(18) &$32^3\times 96$    &-0.2295 & -0.2050 & 303.2(1.3) &524.6(1.8)   &567 \\
    F48P30      & 6.41   &0.07746(18)  & $48^3\times 96$ &-0.2295 & -0.2050   &  303.4(0.9) & 523.6(1.4)  &201\\
\end{tabular}
\end{ruledtabular}
\caption{Parameters of the ensembles. The listed parameters are the coupling $\beta$,  the lattice spacing $a$, the volume $(L/a)^3\times T/a$,  the bare quark masses for the light($am_l$) and strange($am_s$) quarks,  the pion/Kaon mass $m_{\pi / K}$ and the number of configurations $N_{conf}$.}
\label{tab:ensemble}
\end{table}

The distillation quark smearing method ~\cite{HadronSpectrum:2009krc} is used to compute the quark propagators. The smearing operator is composed of a small number($N_{ev}$) of the eigenvectors associated with the $N_{ev}$ lowest eigenvalues of the three-dimensional Laplacian defined in terms of the HYP-smeared gauge field. The number of eigenvectors $N_{ev}$ is 100 for the ensemble F32P30 and 200 for the ensemble F48P30.  

\section{Spectrum determination}
We are interested in the $\Lambda_c \Lambda_c$ scattering with $I(J^P) = 0(0^+)$. The coupled channels $\Xi_{cc} N$ and $\Sigma_c \Sigma_c$ will also be investigated. In this section, we present the spectrum of the relevant single particles, i.e.,  $\Lambda_c$, $\Xi_{cc}$, $N$ and $\Sigma_c$. Then we discuss the finite-volume spectrum of the two-particle systems, which will be used to extract the scattering parameters through L\"uscher's method. 
\label{sec:Spectrum}
\subsection{Single-baryon spectrum}
The interpolating operators of $\Lambda_c$, $\Xi_{cc}$, $N$ and $\Sigma_c$ can be expressed as
\ba
\Lambda_{c, \alpha} &=& \epsilon^{ijk}(u^{i T} C\gamma_5 d^j ) c^k_\alpha, \\
\Xi_{cc,\alpha}^{++} &=& \epsilon^{ijk}(u^{i T} C\gamma_5 c^j ) c^k_\alpha, \quad \Xi_{cc, \alpha}^{+} = \epsilon^{ijk}(d^{i T} C\gamma_5 c^j ) c^k_\alpha, \\
p_\alpha &=& \epsilon^{ijk}(u^{i T} C\gamma_5 d^j ) u^k_\alpha,  \quad n_\alpha = \epsilon^{ijk}(u^{i T} C\gamma_5 d^j ) d^k_\alpha, \\
\Sigma_{c, \alpha}^{++} &=& \epsilon^{ijk}(u^{i T} C\gamma_5 c^j ) u^k_\alpha, \quad \Sigma_{c, \alpha}^{+} = \frac{1}{\sqrt{2}}\epsilon^{ijk}[(u^{i T} C\gamma_5 c^j) d^k_\alpha + (d^{i T} C\gamma_5 c^j ) u^k_\alpha] , \quad \Sigma_{c, \alpha}^{0} = \epsilon^{ijk}(d^{i T} C\gamma_5 c^j ) d^k_\alpha ,
\ea
where $u, d, c$ represent the quark fields, $i, j, k$ are color indices and $\alpha = \{1,2,3,4\}$ is the Dirac four-spinor index in the Dirac basis. These single-particle operators will be the building blocks of the two-baryon operators.  

\begin{figure}
\includegraphics[width =0.33 \textwidth]{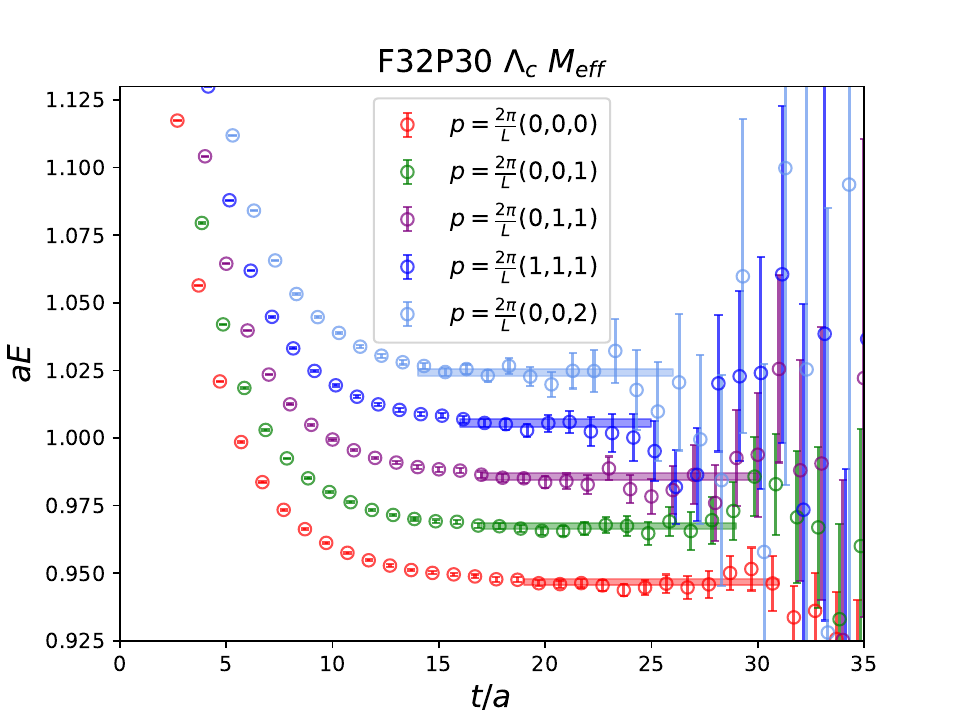}\includegraphics[width =0.33 \textwidth]{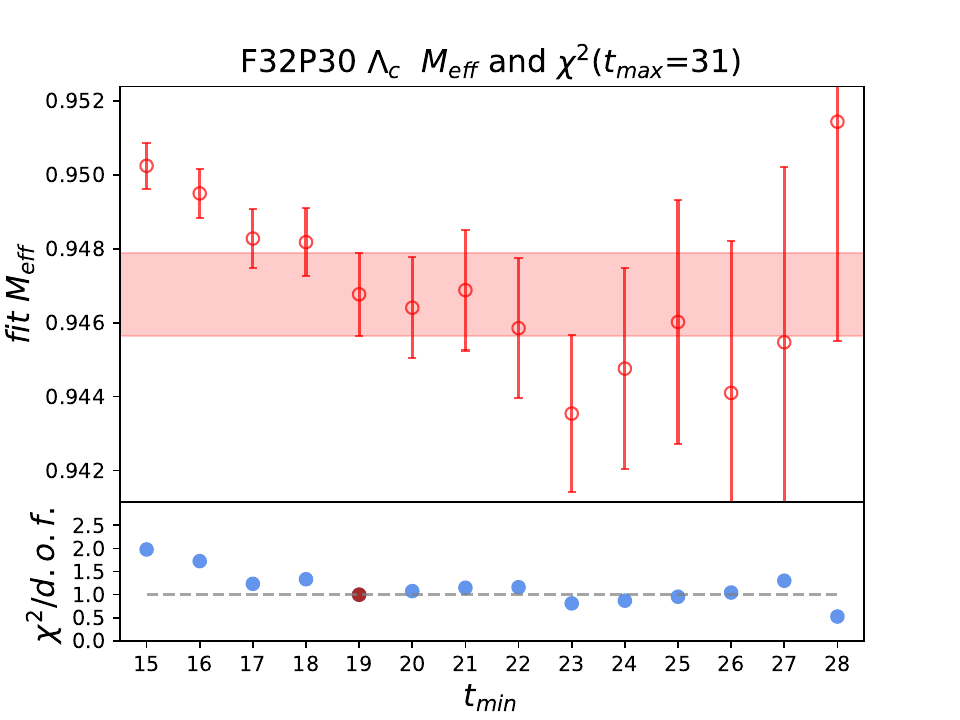}\includegraphics[width =0.33 \textwidth]{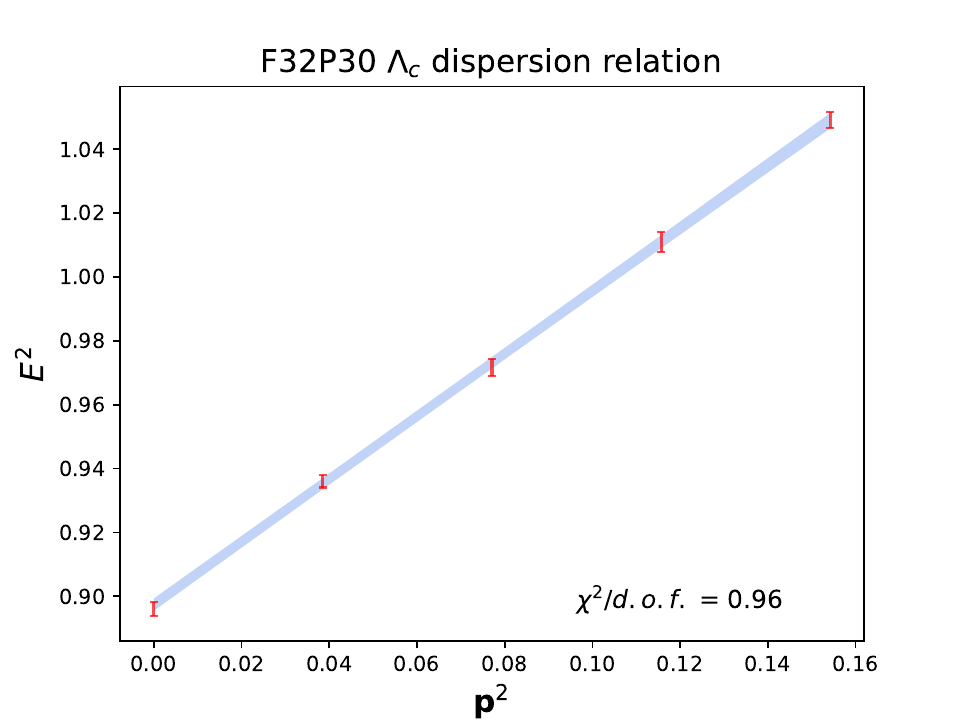}
\caption{Left: the effective mass of $\Lambda_c$ at the five momenta. The fitted mass and fitting range are indicated by the horizontal band.  Middle: The fitted mass of the zero momentum $\Lambda_c$ at different $t_{min}$. The corresponding values of $\chi^2/ d.o.f$ of the fits are also shown in the lower part of the plot, where the chosen $t_{min}$ is highlighted by the red point. Right: The fit of the dispersion relation for $\Lambda_c$. All of the three plots are for the ensemble F32P30.}
\label{fig:SingleParticle}
\end{figure}

The masses of the baryons are then obtained from the correlation functions of the above interpolating operators 
\be
C(\mathbf{p}, t) = \sum_{t_{src}} \langle 0 | \mathcal{O}_{\alpha} (\mathbf{p}, t+t_{src})P^+_{\alpha\beta} \mathcal{O}^\dagger_{\beta} (\mathbf{p}, t_{src}) |0 \rangle  ,
\ee
where $P^+ = \frac{1}{2}(1+\gamma_4)$ is the positive parity projection operator, the source-time $t_{src}$ is summed over all time slices to increase the statistics, $\mathcal{O}(\mathbf{p}, t)$ is the momentum projected operator defined as $\mathcal{O}(\mathbf{p}, t) = \sum_{\mathbf{x}} e^{-i\mathbf{p}\mathbf{x}} \mathcal{O}(\mathbf{x}, t) $. The dispersion relation $E^2 = m_0^2 + c^2\mathbf{p}^2$  is investigated by calculating the single-particle energy at the five lowest momenta on lattice: $\mathbf{p} =$ $(0,0,0)$, $(0,0,1)$, $(0,1,1)$, $(1,1,1)$, $(0,0,2)$ in units of $\frac{2\pi}{L}$. The effective mass of $\Lambda_c$ at the five momenta for the ensemble F32P30 is shown in the left panel of Fig.\ref{fig:SingleParticle}. We fit the correlation functions to an exponential form $C(t) = Ae^{-Et}$ to obtain the energies. The range for fitting, denoted as $[t_{min}, t_{max}]$ is selected as follows: $t_{max}$  is set to a sufficiently large value where the error becomes significant. We then adjust the $t_{min}$ until the fitted mass stabilizes, indicating that the fitting is reliable. The $\chi^2/ d.o.f$ is generally around 1. In the middle panel of Fig.\ref{fig:SingleParticle}, we display the fitted mass at different $t_{min}$, along with the corresponding $\chi^2/ d.o.f$ for fitting the zero momentum $\Lambda_c$ correlation function. The chosen $t_{min}$ is highlighted by the dark red point in the lower panel. For each baryon, we fit the five energies at the five momenta to the dispersion relation to get the parameters $m_0$ and $c$. The results are collected in TABLE~\ref{tab:SingleParticleSpectra}. The values of $c$ for $\Lambda_c$, $\Sigma_c$ and $\Xi_{cc}$ tend to deviate from 1, primarily due to lattice artifacts stemming from the charm quark. The effects of these lattice artifacts will be discussed in more detail later on. In the right panel of Fig.\ref{fig:SingleParticle}, we display the fitting of the dispersion relation for $\Lambda_c$. 

\begin{table*}[tb]
\begin{tabular}{|c|c|c|c|c|c|c|c|c|}
\hline
  \multirow{2}{*}{} & \multicolumn{2}{c|}{$\Lambda_c$} & \multicolumn{2}{c|}{$\Sigma_c $} &\multicolumn{2}{c|}{$ \Xi_{cc}$}  &\multicolumn{2}{c|}{$N$} \\
  \cline{2-9} 
   &$m_0$(GeV) &$c$  &$m_0$(GeV)  &$c$  &$m_0$(GeV)  & $c$  &$m_0$(GeV)  & $c$ \\
 \hline
F32P30 &2.413(3) &0.991(8) & 2.572(3)&1.01(1) &3.747(1) &0.948(5)&1.070(4)&1.01(1)  \\
 \hline
 F48P30 &2.410(1) & 0.988(7)& 2.566(1)& 1.01(1)&3.7504(7) &0.931(8) & 1.062(2)&1.005(8) \\
 \hline
\end{tabular}
\caption{Fit results of the dispersion relation for $\Lambda_c$, $\Sigma_c$,  $\Xi_{cc}$ and $N$. }
\label{tab:SingleParticleSpectra}
\end{table*}

\subsection{Spectra of dibaryon systems}
We will focus on the S-wave scattering in the $I(J^P) = 0(0^+)$ channel. Therefore we construct the dibaryon operators in the $A_1^+$ irreducible representation(irrep) of the octahedral group($O_h$), which is the rotational symmetry group on lattice. The operators for $\Lambda_c\Lambda_c$, $\Xi_{cc}N$ and $\Sigma_c\Sigma_c$ are:
\ba
\mathcal{O}_{\Lambda_c\Lambda_c}({|\mathbf{p}|,t)}  &=& \sum_{\alpha, \beta, \mathbf{p}} c_{\alpha, \beta, \mathbf{p}} \Lambda_{c, \alpha}(\mathbf{p},t) \Lambda_{c, \beta}(-\mathbf{p},t), \\
\mathcal{O}_{\Xi_{cc}N}(|\mathbf{p}|,t)) &= &  \sum_{\alpha, \beta, \mathbf{p}} c_{\alpha, \beta, \mathbf{p}} \big(\Xi_{cc, \alpha}^{++}(\mathbf{p},t) n_{\beta}(-\mathbf{p},t) - \Xi_{cc, \alpha}^{+}(\mathbf{p},t) p_{\beta}(-\mathbf{p},t)\big), \\
\mathcal{O}_{\Sigma_c\Sigma_c} (|\mathbf{p}|,t))  &=&  \sum_{\alpha, \beta, \mathbf{p}} c_{\alpha, \beta, \mathbf{p}} \big(\Sigma_{c, \alpha}^{++}(\mathbf{p},t) \Sigma_{c,\beta}^{0}(-\mathbf{p},t) - \Sigma_{c, \alpha}^{+}(\mathbf{p},t) \Sigma_{c,\beta}^{+}(-\mathbf{p},t) + \Sigma_{c, \alpha}^{0}(\mathbf{p},t) \Sigma_{c,\beta}^{++}(-\mathbf{p},t) \big), 
\ea
where the coefficients $c_{\alpha, \beta, \mathbf{p}}$ are chosen so that the operators transform in the $A_1^+$ irrep. To be specific, for a given $|\mathbf{p}|$, the non-zero coefficients are $c_{1, 2, \mathbf{p}} = 1$ and $c_{2, 1, \mathbf{p}} = -1$ for all $\mathbf{p}$. We will only use the operators with zero total momentum. 

The spectra of the dibaryon systems in finite volume is determined from the matrix of correlation functions of the operators: 

\be
C_{ij}(t) = \sum_{t_{src}} \langle 0 |\mathcal{O}_i(t+t_{src}) \mathcal{O}_j^\dagger(t_{src})|0\rangle.
\ee

Solving the generalized eigenvalue problem(GEVP)
\be
C(t) v^n(t) = \lambda^n(t) C(t_0) v^n(t),
\ee
the energies can be extracted from the time dependence of the eigenvalues $\lambda^n(t)$. We set $t_0=4$  and fit the eigenvalues to a two-exponential form $\lambda^n(t) = A_n e^{-E_n(t-t_0)} + (1-A_n) e^{-E_n^\prime(t-t_0)}$ to get the $n$-th energy level $E_n$. 
The overlap factor between the $n$-th GEVP eigenstate and $i$-th operator can be evaluated as $\langle n|\mathcal{O}_i|0\rangle = \sqrt{2 m_n} v^{n*}_j C_{ji}(t_0)$\cite{Dudek:2010wm}.

\begin{figure}
\begin{tabular}{cc}
\includegraphics[width =0.6 \textwidth]{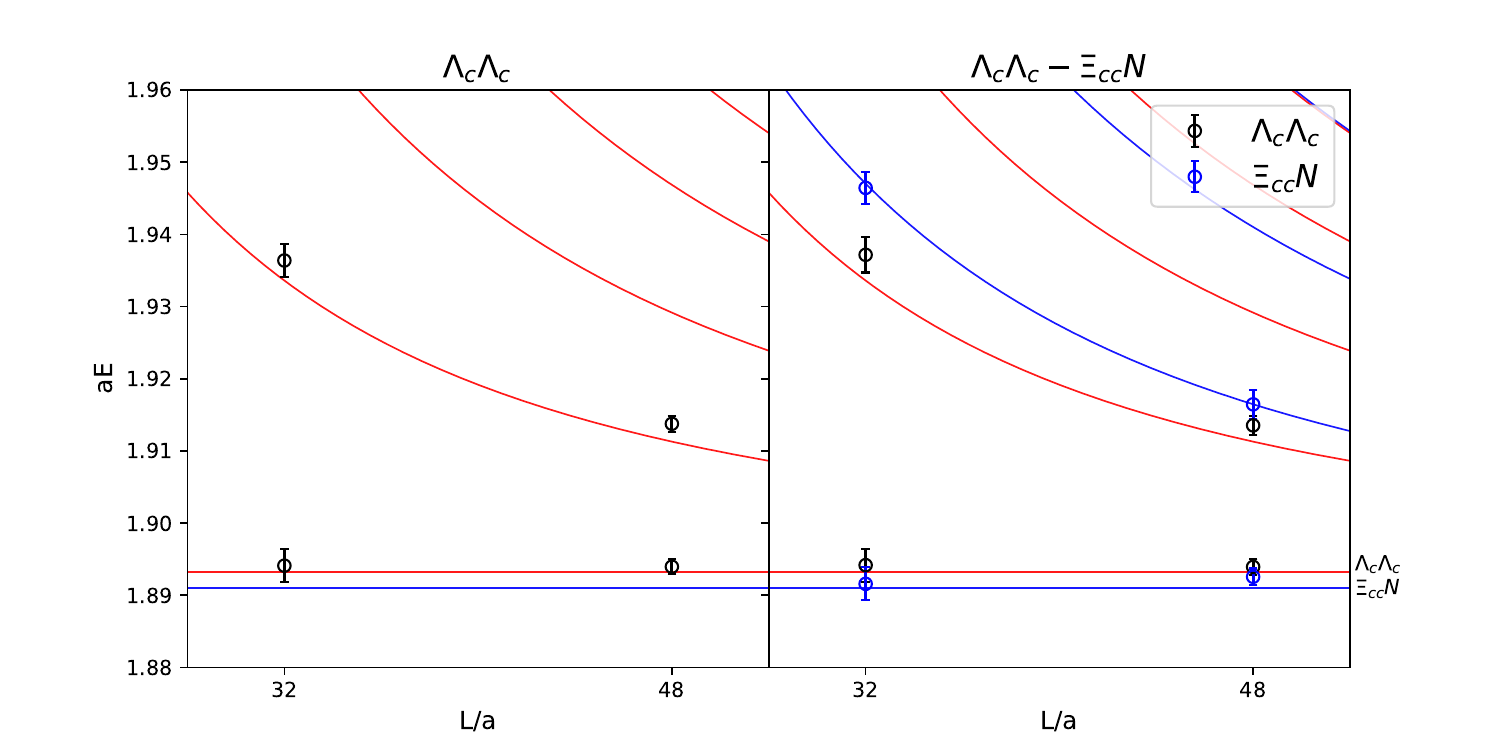}&\includegraphics[width =0.37 \textwidth]{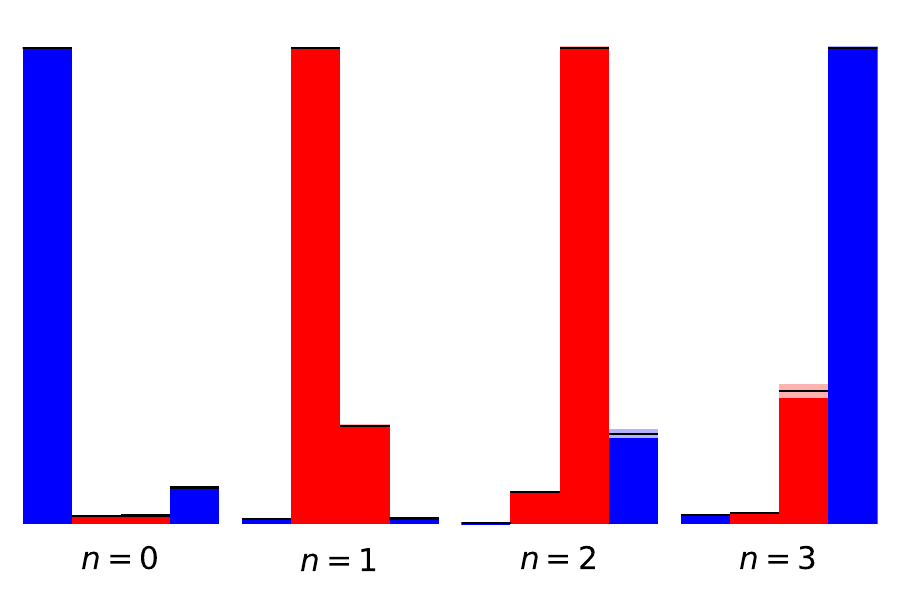} \\
(a) & (b)
\end{tabular}
\caption{(a) Comparison of the energies using both $\Lambda_c\Lambda_c$ and $\Xi_{cc}N$ operators(right), and using only $\Lambda_c\Lambda_c$ operators(left) for the ensembles F32P30 and F48P30. The red and blue lines represent non-interacting $\Lambda_c\Lambda_c$ and $\Xi_{cc}N$ channel threshold, respectively. In the right panel, the black and blue points represent the energy levels predominantly overlap with the $\Lambda_c\Lambda_c$ and $\Xi_{cc}N$ operators, respectively. (b) The overlaps of the operators onto the eigenstates from the GEVP analysis using the $\Lambda_c\Lambda_c$ and $\Xi_{cc}N$ operators for the ensemble F32P30. The red and blue bars represent the operators $\Lambda_c\Lambda_c$ and $\Xi_{cc}N$ respectively. $n=0,1,2,3$ are the eigenstates with energy from low to high.}
\label{fig:XiccNcoupling}
\end{figure}

In order to investigate the coupling between $\Lambda_c\Lambda_c$ and $\Xi_{cc} N$, we compute the matrix of the correlation functions of the four operators: $\mathcal{O}_{\Lambda_c\Lambda_c}({\mathbf{p}^2=0,1)}$ and $\mathcal{O}_{\Xi_{cc}N}(\mathbf{p}^2=0,1))$. We found negligible coupling between the $\Lambda_c\Lambda_c$ and $\Xi_{cc}N$ operators. In Fig.~\ref{fig:XiccNcoupling}(a), we compare the energy levels obtained from the GEVP analysis using the four operators(right panel) and those obtained using only the two $\Lambda_c\Lambda_c$ operators(left panel). In the right panel, the black and blue points present the energy levels predominantly overlap with the $\Lambda_c\Lambda_c$ and $\Xi_{cc}N$ operators, respectively. The overlaps of the operators onto the eigenstates are shown in Fig.~\ref{fig:XiccNcoupling}(b) for the ensemble F32P30. The overlaps for the ensemble F48P30 exhibit a similar pattern. These overlaps are calculated at time slice $t=14$, which is the region where the fit windows are typically situated.
It is evident that the two energy levels primarily associated with the $\Lambda_c\Lambda_c$ operators are almost identical to the energies obtained using only the $\Lambda_c\Lambda_c$ operators.  Therefore, in this study we will not consider the coupling effects from the $\Xi_{cc}N$ channel. 

\begin{figure}
\includegraphics[width =0.3 \textwidth]{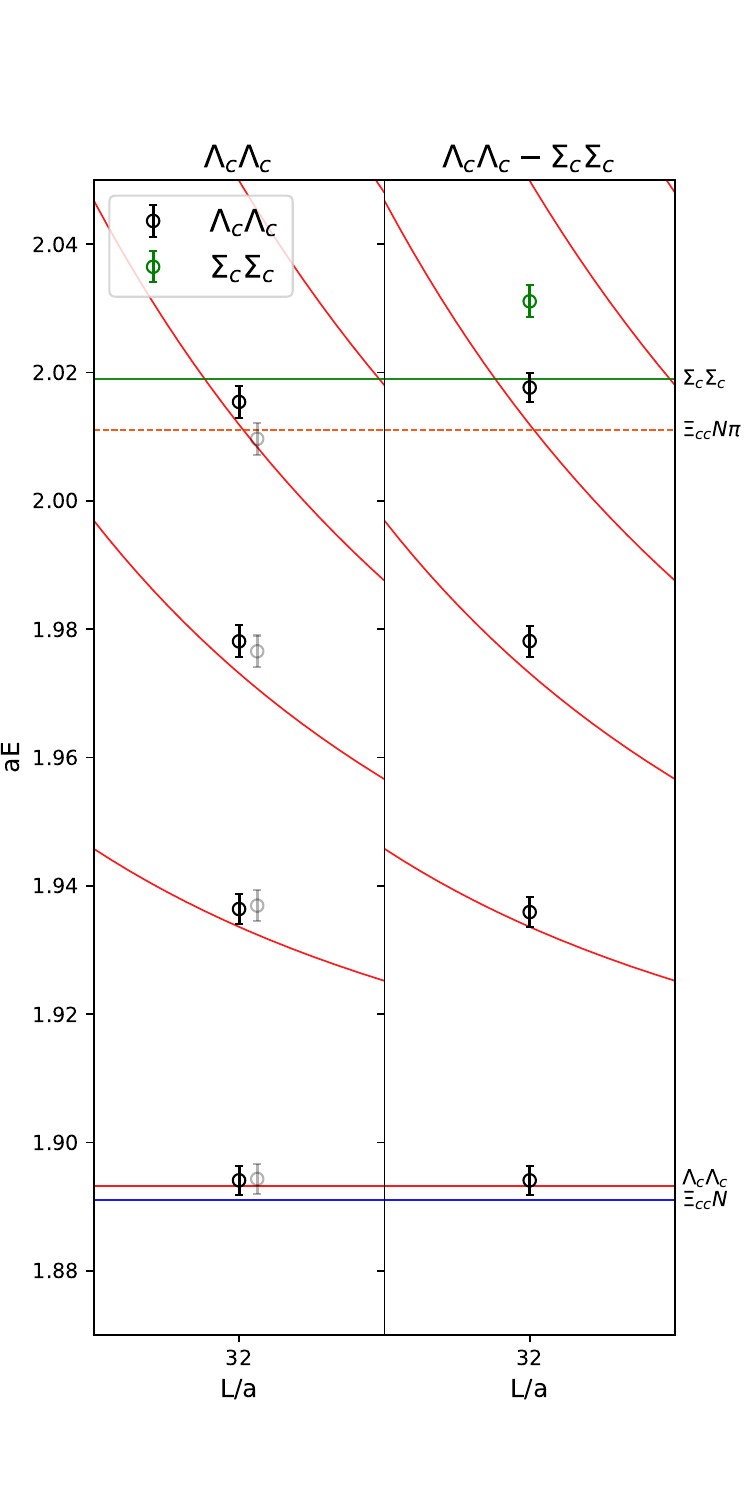}
\caption{Comparison of the energies using both $\Lambda_c\Lambda_c$ and $\Sigma_{c}\Sigma_{c}$ operators(right), and using only $\Lambda_c\Lambda_c$ operators(left) for the ensemble F32P30. The black points in the left panel are the results from the GEVP analysis, while the gray points are obtained from the diagonal matrix elements. In the right panel, the black and green points represent the energy levels predominantly overlap with the $\Lambda_c\Lambda_c$ and $\Sigma_c\Sigma_c$ operators, respectively. The red lines are the free energies of $\Lambda_c\Lambda_c$. The thresholds of $\Xi_{cc}N$, $\Xi_{cc}N\pi$ and $\Sigma_c\Sigma_c$ are also shown by the blue, orange and green lines, respectively.}
\label{fig:SigcSigccoupling}
\end{figure}

$\Sigma_c\Sigma_c$ can also couple to $\Lambda_c\Lambda_c$. Its threshold is much higher than $\Lambda_c\Lambda_c$. As will shown later, the energy range in which we perform scattering analysis lies well below the $\Sigma_c\Sigma_c$ threshold. However, we still checked the effects of $\Sigma_c\Sigma_c$ channel by computing the correlation function matrix of the five operators: $\mathcal{O}_{\Lambda_c\Lambda_c}({\mathbf{p}^2=0,1,2,3)}$ and $\mathcal{O}_{\Sigma_c\Sigma_c}({\mathbf{p}^2=0)}$ for the ensemble F32P30. In Fig.~\ref{fig:SigcSigccoupling}, we compare the energy levels from the GEVP analysis with and without  $\Sigma_{c}\Sigma_{c}$ operator. The lowest three energies, which predominantly couple to the operators $\mathcal{O}_{\Lambda_c\Lambda_c}({\mathbf{p}^2=0,1,2)}$, exhibit close agreement between the two cases. The energy level close to the $\Lambda_c\Lambda_c({\mathbf{p}^2=3)}$ free energy is shifted slightly upon the inclusion of the $\Sigma_c\Sigma_c$ operator. 
For the subsequent scattering analysis, we cut the energy at around $aE =  1.98$. In this range, the $\Lambda_c\Lambda_c$ scattering should not be affected by the coupling from $\Sigma_c\Sigma_c$. To explore the coupled channel scattering at the energy range close to the $\Sigma_c\Sigma_c$ threshold, additional energy levels in this range would be required by incorporating operators with higher momenta. Moreover, the inclusion of the $\Xi_{cc}N\pi$ three-body system would be necessary, which is out of the scope of this study. 

In the remaining of this paper, we will focus on the single channel $\Lambda_c\Lambda_c$ scattering. For the ensemble F32P30, we computed the correlation function matrix of the four operators $\Lambda_c\Lambda_c({\mathbf{p}^2=0,1,2,3)}$ and extracted four energy levels from the GEVP analysis. The highest one is not included in the scattering analysis since it is close to the $\Sigma_c\Sigma_c$ and $\Xi_{cc}N\pi$ thresholds. We found that the off-diagonal elements of correlation function matrix are very small and does not contribute to the determination of the energy levels. The energy levels obtained from the diagonal matrix elements are presented in the left panel of Fig.~\ref{fig:SigcSigccoupling}, alongside those obtained from GEVP method. The results are nearly identical except the highest one of which the signal is pretty noisy and is excluded in the scattering analysis due to its closeness to the $\Sigma_c\Sigma_c$ and $\Xi_{cc}N\pi$ thresholds. For the ensemble F48P30, we employ five operators $\Lambda_c\Lambda_c({\mathbf{p}^2=0,1,2,3,4)}$.  To save computational cost, only the diagonal correlation functions are calculated. 

\begin{figure}
\includegraphics[width =0.2 \textwidth]{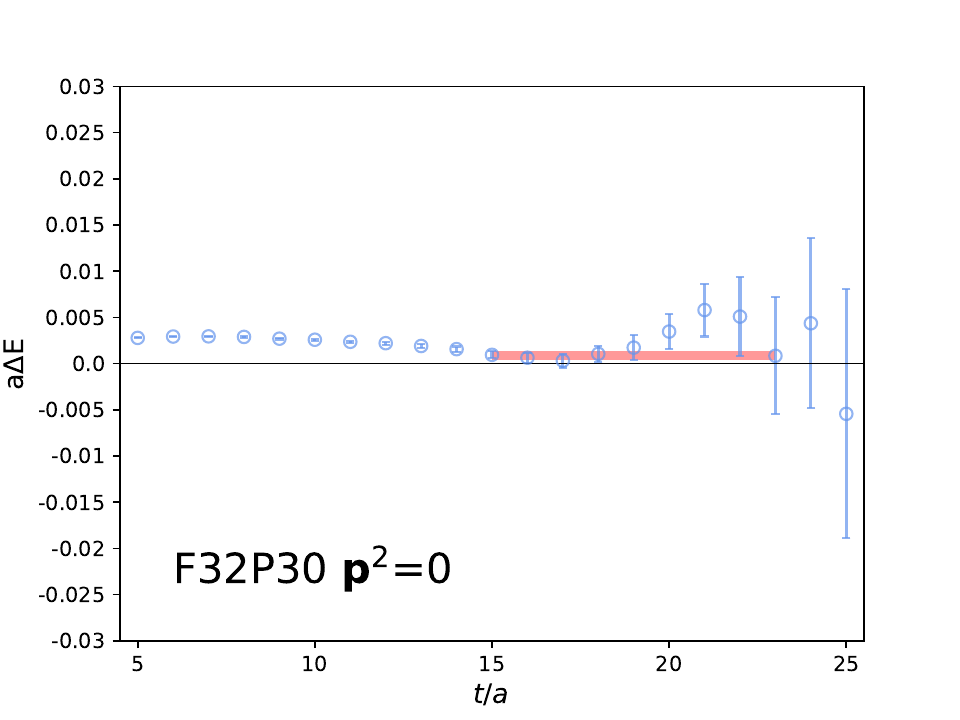}\includegraphics[width =0.2 \textwidth]{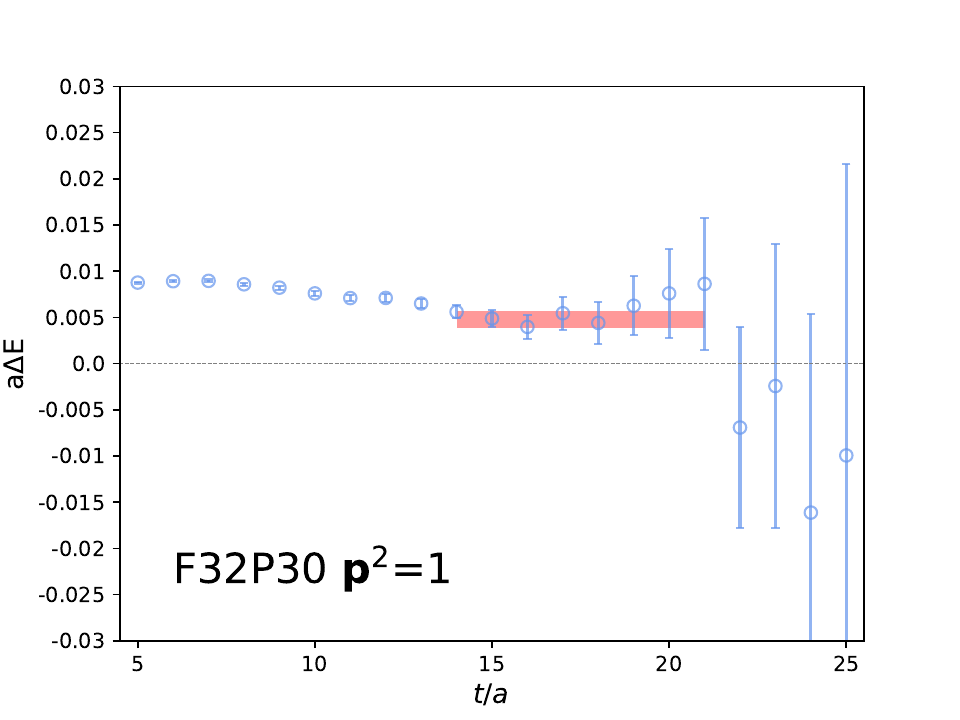}\includegraphics[width =0.2 \textwidth]{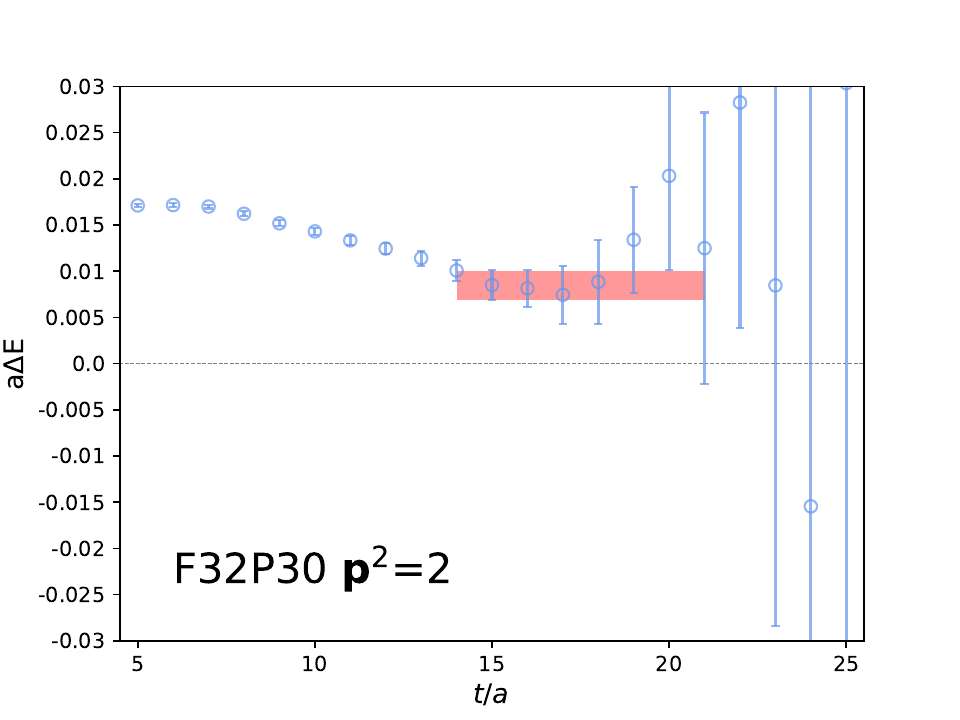}\includegraphics[width =0.2 \textwidth]{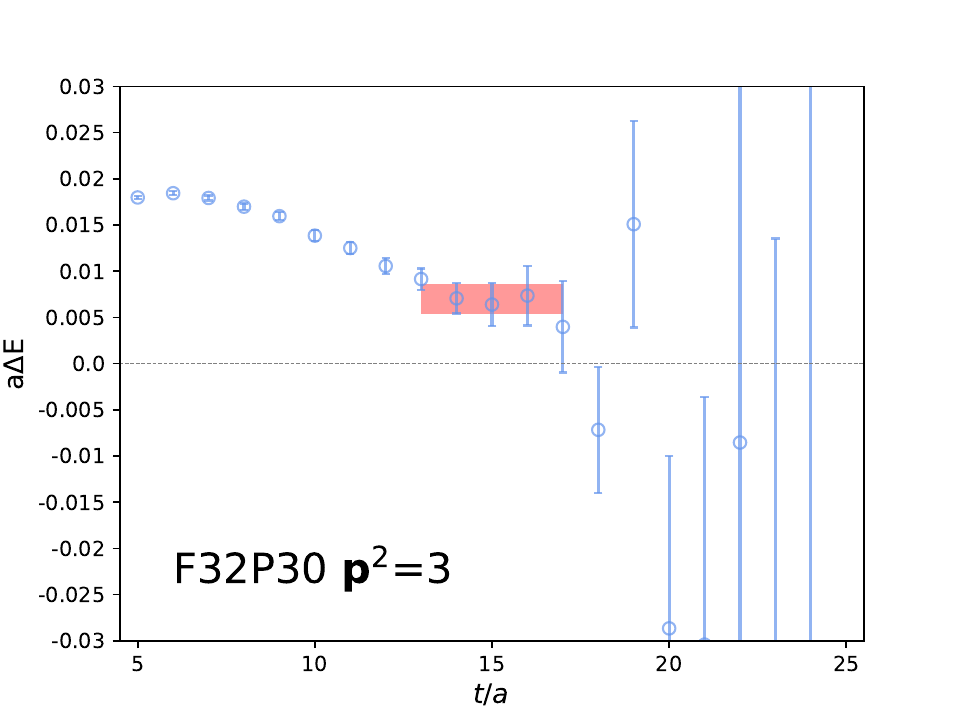}
\includegraphics[width =0.2 \textwidth]{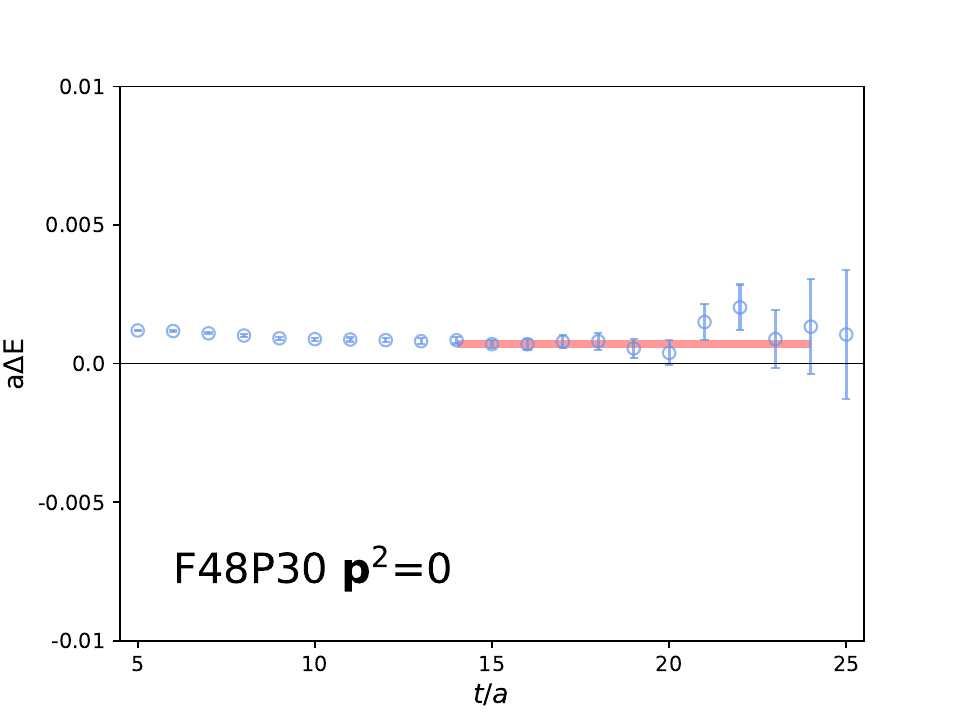}\includegraphics[width =0.2 \textwidth]{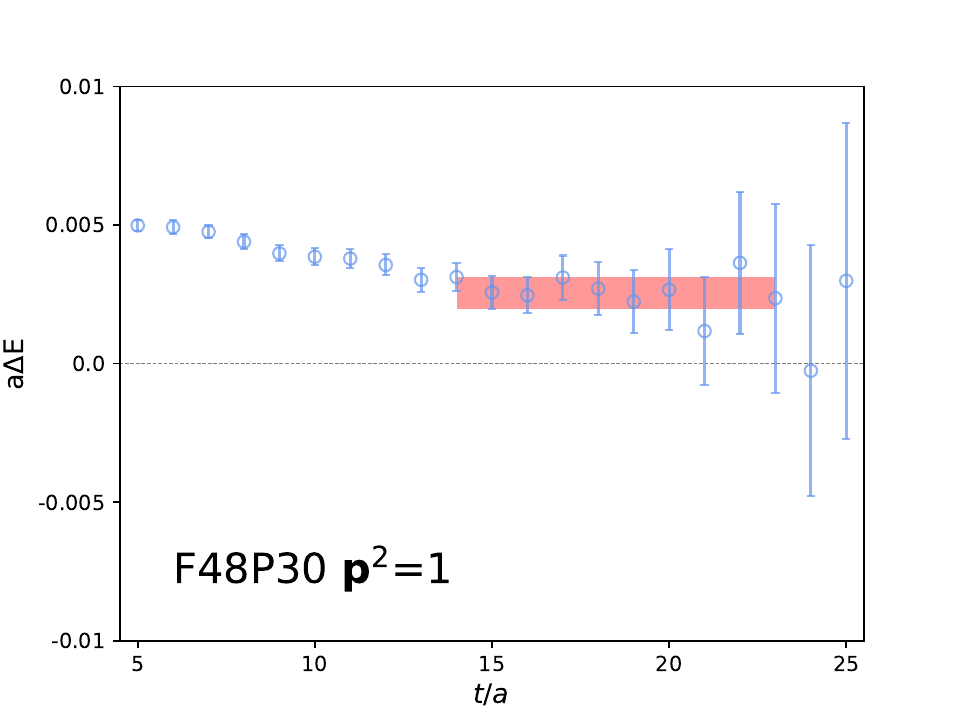}\includegraphics[width =0.2 \textwidth]{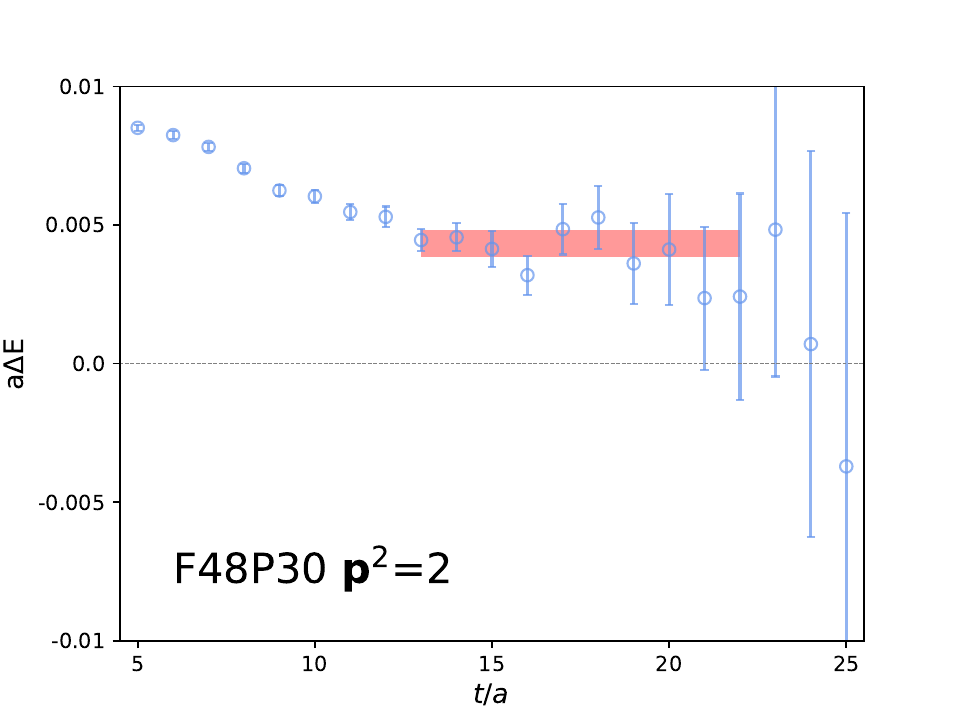}\includegraphics[width =0.2 \textwidth]{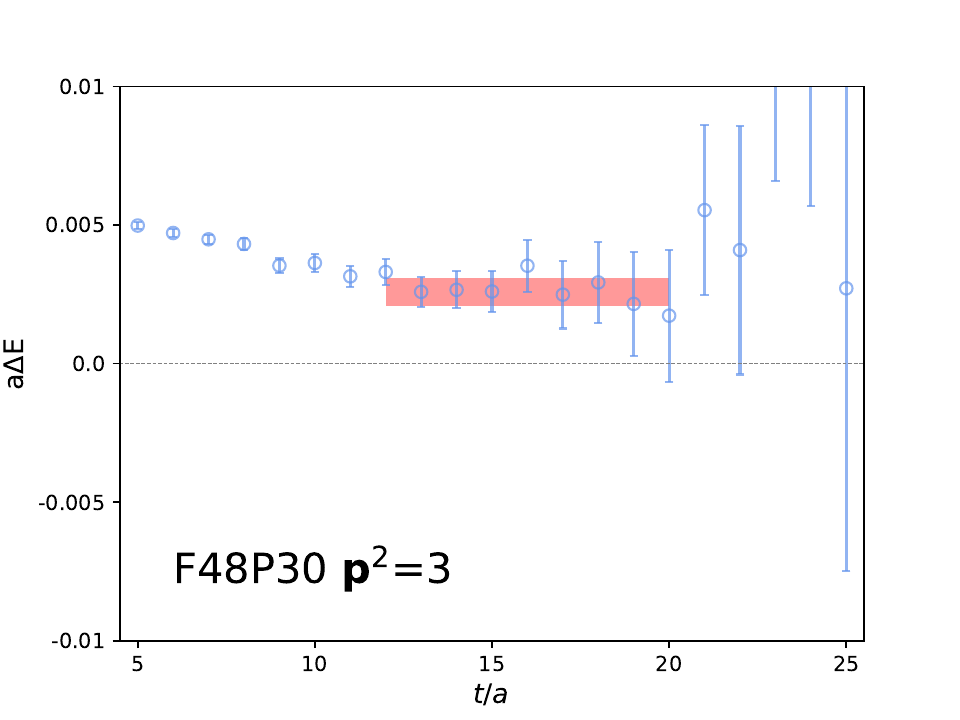}\includegraphics[width =0.2 \textwidth]{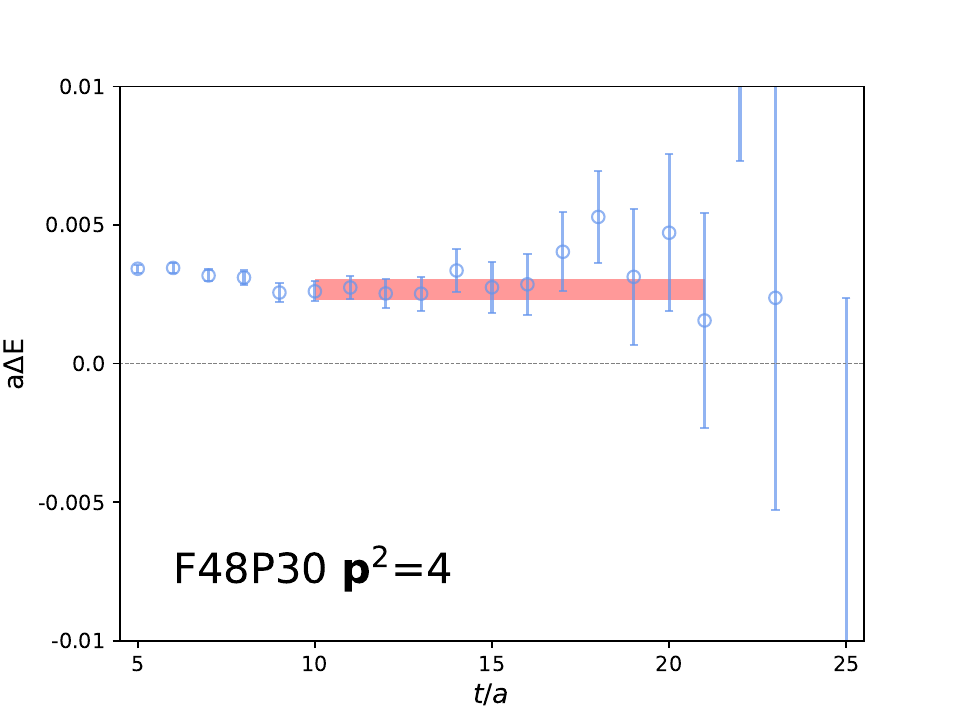}
\caption{Effective mass calculated from the ratio defined in Eq.~\ref{eq:ratio}. The red horizontal bands indicate the fitted values of $\Delta E$ and the fitting ranges.}
\label{fig:diLambdaDE}
\end{figure}

\begin{figure}
\includegraphics[width =0.3 \textwidth]{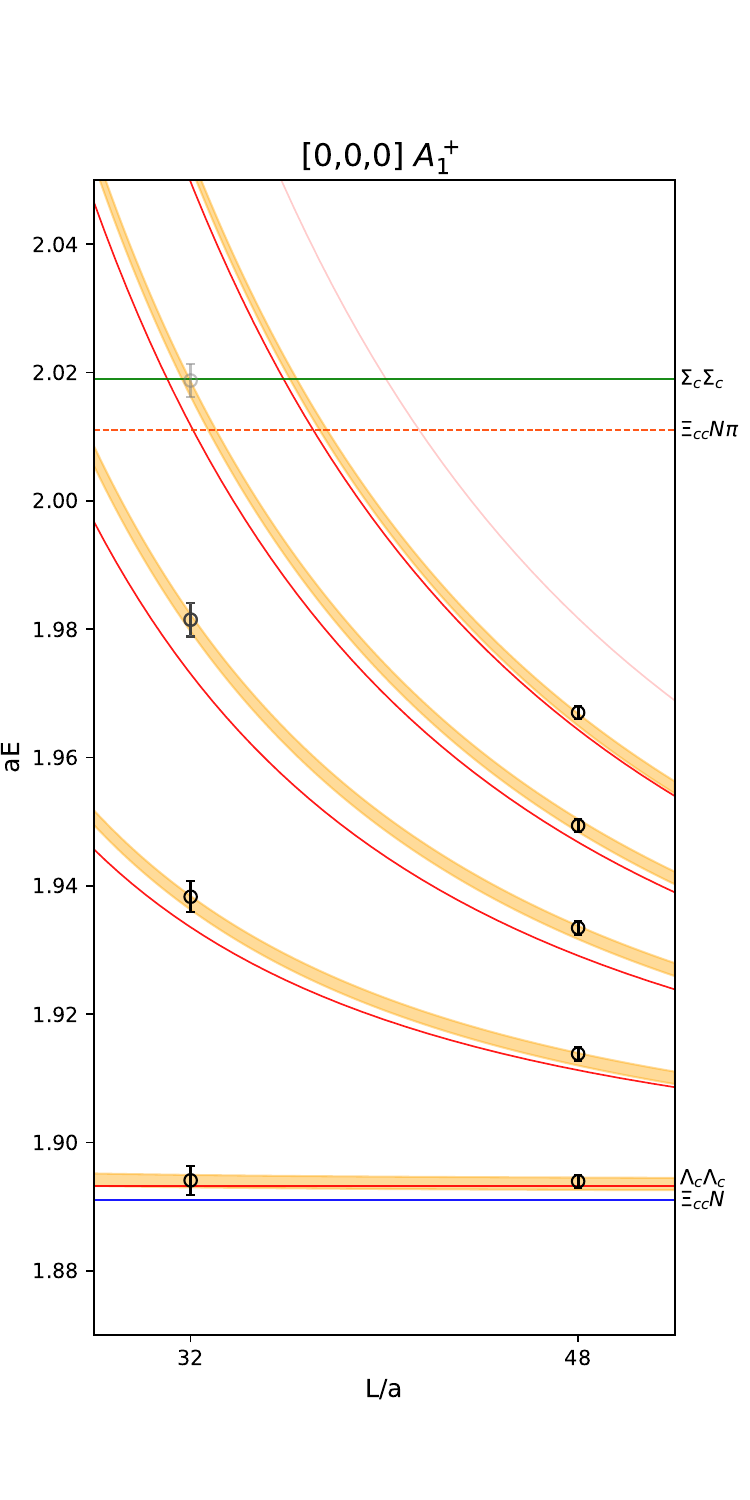}
\caption{Finite-volume spectrum of $\Lambda_c\Lambda_c$ system. The data points are the energy levels calculated from Eq.~\ref{eq:En}. The red solid lines are the free energies of $\Lambda_c\Lambda_c$. The blue, green and red dashed lines are the thresholds of $\Xi_{cc} N$, $\Sigma_c \Sigma_c$ and $\Xi_{cc} N\pi $, respectively. The orange bands are the solutions of the L\"uscher's equation from ``fit1", which will be explained in the next section.  
}
\label{fig:LamcLamcSpectra}
\end{figure}

In order to accurately extract the energy levels, it is favorable to fit the ratio of the two-baryon correlation function to the square of the single baryon correlation function:
\be
R(t) = \frac{C^n_{\Lambda_c\Lambda_c}(t)}{C^n_{\Lambda_c}(t) C^n_{\Lambda_c}(t)} \sim Ae^{-\Delta E_n t}, \quad n=0,1,2,\cdots,
\label{eq:ratio}
\ee
where $C^n_{\Lambda_c\Lambda_c}(t)$ is the $n$-th eigenvalue from the GEVP analysis for the ensemble F32P30 or the diagonal $\Lambda_c\Lambda_c$ correlation functions in the case of F48P30, $C^n_{\Lambda_c}(t)$ is the correlation function of the single $\Lambda_c$ with momentum $\mathbf{p}^2 = n$. $\Delta E_n$, representing the energy shift of the $\Lambda_c\Lambda_c$ system with respect to two free $\Lambda_c$ with momentum $\mathbf{p}^2 = n$, is obtained by fitting $R(t)$ to an exponential function. In Fig.~\ref{fig:diLambdaDE}, we display the effective mass calculated from the ratio $R(t)$ for all energy levels, the fitted $\Delta E$ and fitting ranges are also illustrated by the horizontal bands in the plot.  The results are collected in TABLE~\ref{tab:diLambdaDE}. The interacting energies of the di-$\Lambda_c$ system are then calculated as:
\be
E_n = \Delta E_n + 2 \sqrt{m^2_{\Lambda_c} + n (\frac{2\pi}{L})^2},
\quad n=0,1,2,\cdots,
\label{eq:En}
\ee
Instead of using the dispersion relation determined by fitting the $\Lambda_c$ energies computed on lattice, where the speed of light $c$ deviates from 1, here we employ the continuum dispersion to estimate the free energies. The $E_n$'s calculated in this way approximate the interacting energies under the continuum dispersion relation, and will be used to determine the scattering parameters through L\"uscher's formula in the subsequent analysis. Since the continuum dispersion relation is implicitly applied in the derivation of L\"uscher's formula and its generalizations, this approach is expected to alleviate the effects of the deviation from the continuum dispersion relation, as has been discussed and applied in the charmed meson scattering calculations ~\cite{Padmanath:2022cvl, Prelovsek:2020eiw, Piemonte:2019cbi}. The $E_n$'s are plotted in Fig.~\ref{fig:LamcLamcSpectra} along with the free energies and the solution of the L\"uscher's equation, which will be explained in the next section.

\begin{table*}[tb]
\begin{tabular}{|c|c|c|c|c|c|}
\hline
& $a\Delta E_0$ & $a\Delta E_1$ & $a\Delta E_2$ & $a\Delta E_3$ & $a\Delta E_4$ \\
  \hline
 F32P30 &0.00085(47) &0.00475(93) &0.0085(15) &0.0070(16) &   \\
 \hline
 F48P30 &0.00070(15) &0.00255(56)  &0.00433(48) &0.00259(50) &0.00268(37)  \\
 \hline
\end{tabular}
\caption{The values of $\Delta E$ fitted from the ratio defined in Eq.~\ref{eq:ratio}.}
\label{tab:diLambdaDE}
\end{table*}

\section{scattering analyses and results}
\label{sec:Results}
L\"uscher's finite volume method provides a direct relation between the energy of a two-particle system in a finite box and the scattering phase shift of the two particles in the infinite volume. We consider two $\Lambda_c$ particles in the rest frame. The finite volume energies are computed in the $A_1^+$ irrep of the $O_h$ group. If we ignore the contributions of the partial waves $l \ge 4$, the L\"uscher's formula reads 
\be
k\cot\delta(k) = \frac{2}{\sqrt{\pi} L} \mathcal{Z}_{00}(1; q^2),
\label{eq:LuscherFormula}
\ee
where $\delta(k)$ is the s-wave scattering phase shift,  the momentum $k$ is related to the finite-volume energy $E$ by $E = 2 \sqrt{m_{\Lambda_c}^2 + k^2}$, $\mathcal{Z}_{00}$ is the zeta function and the dimensionless variable $q = \frac{L}{2\pi} k$. 

We use the effective range expansion up to $\mathcal{O}(k^2)$ to parameterize the phase shift:
\be
k\cot\delta(k) = \frac{1}{a_0} + \frac{1}{2} r_0 k^2. 
\label{eq:ERE}
\ee
The parameters $a_0$ and $r_0$ is determined by minimizing the $\chi^2$ defined as 
\begin{align}
\chi^2 =  \sum_{L,n,n^\prime} [ E_n(L) - E_n^{sol.}(L,  a_0, r_0) ] C^{-1}_{nn^\prime} [ E_{n^\prime}(L) - E_{n^\prime}^{sol.}(L, a_0, r_0) ] ,
\label{Eq:chi2}
\end{align}
where $E_n(L)$ is the $n$-th energy level obtained on the lattice with size $L$, $E_n^{sol.}(L, a_0, r_0)$ is the $n$-th solution of Eq.~\ref{eq:LuscherFormula} with parameters $a_0$ and $r_0$. $C$ is the covariance matrix of $E_n(L)$. 

In order to check the finite volume effects, we determined the scattering parameters using the energies from the ensembles F32P30 and F48P30 separately. The results are consistent with each other within 1$\sigma$ of the statistical uncertainty, suggesting negligible finite volume effects. Our final results are then determined using the energies from both ensembles. Considering that the effective range expansion is valid only near the threshold, we also estimated the systematic error arising from variations of $k^2$ range in the expansion. We performed the fit using two different data sets: 1. using all energy levels of the two ensembles, and 2. exclude the highest two energy levels. The discrepancy between these two fits is taken as the systematic error from the ERE parameterization. The fit results are summarized in Table~\ref{tab:ScattFit}. Since we only have one lattice spacing, we are not able to estimate the systematic uncertainty associated with the finite lattice spacing. However, in the dispersion relation of $\Lambda_c$, the value of $c^2$ is deviated from 1 by around 4\%, we expect that this effect should be much smaller than the statistical error. Further investigation on the discretization effects needs to do the calculations at various lattice spacings. Different choice of the fit range used to determin the $\Delta E$ values from Eq.~\ref{eq:ratio} may also cause variance in the final results. To assess this systematic effect, we select 5-8 different fit ranges for each of the eight energy levels, ensuring that the $\chi^2$ value remain reasonable. Then we randomly chose one fit range for each energy level and determine the scattering parameters from these energy levels. This procedure is repeated 100 times. The mean values and standard errors of the scattering parameters from these 100 measurements are $a_0 = -0.21(4) \text{fm}, r_0 = -0.03(18) \text{fm}$, which are in excellent agreement with the ``fit1" results in Table~\ref{tab:ScattFit}. The errors here are then estimated as the systematic errors arising from the choice of fit ranges. These errors are added quadratically with the systematic error from the ERE parameterization as the total systematic error in our final results.

\begin{figure}
\includegraphics[width=0.6\textwidth]{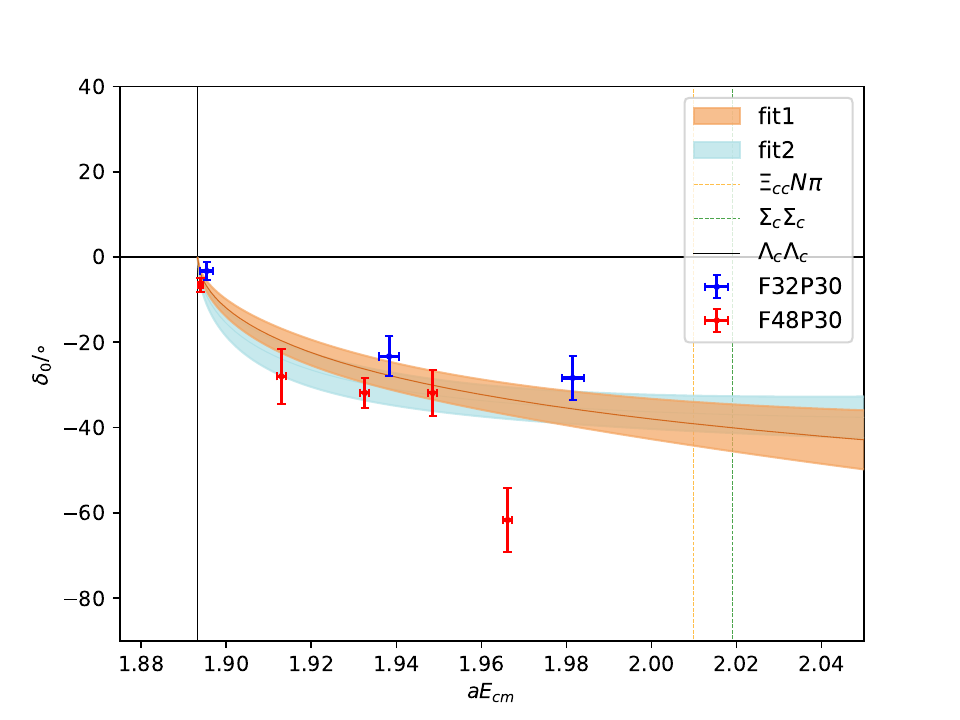}
\caption{Energy dependence of the phase shift. }
\label{fig:PhaseShifts}
\end{figure}

Our final result of the scattering length and effective range are 
\be
a_0 = -0.21(4)(8) \text{ fm}, \quad \quad r_0 = -0.05(13)(25) \text{ fm},
\ee
where the first error is statistical error and the second is systematic error. The energy dependence of the phase shift  is plotted in Fig.~\ref{fig:PhaseShifts}. In the scattering amplitude, there is no poles in  the energy range we investigated. 
\begin{table*}[tb]
\begin{tabular}{|c|c|c|c|c|}
\hline
  & \multirow{2}{*}{F32P30} &\multirow{2}{*}{F48P30} & \multicolumn{2}{c|}{F32P30\&F48P30} \\
  \cline{4-5} 
   & &                                                                              & fit1 & fit2 \\
  \hline
  $a_0/\mathrm{fm}$ &-0.21(5) &-0.21(4) &-0.21(4) &-0.28(6) \\
  \hline
  $r_0/\mathrm{fm}$ &-0.22(21) &0.11(15) &-0.05(13) &-0.23(11) \\
  \hline
  $\chi^2/dof$& 0.08 &2.3 &1.6 &0.3 \\
  \hline
 \end{tabular}
\caption{Results of the scattering parameters by fitting the energies from the ensembles F32P30 and F48P30 separately and collectively(F32P30\&F48P30). ``fit1" uses all energy levels of the two ensembles, while ``fit2" excludes the highest two energy levels.}
\label{tab:ScattFit}
\end{table*}

\section{Summary and discussions}
\label{sec:Summary}
We present the lattice QCD study of the $\Lambda_c\Lambda_c$ scattering based on two gauge ensembles with $2+1$ dynamical flavors at pion mass $ 303$\,MeV and lattice spacing $0.07746$\,fm. The finite volume spectrum of the $\Lambda_c\Lambda_c$ system is calculated in the rest frame and L\"uscher's finite volume formalism is utilized to determine the scattering parameters from the finite-volume spectrum. It is found that the interaction between two $\Lambda_c$ baryons is repulsive. The scattering length and effective range are $a_0 = -0.21(4)(8) \text{ fm}, r_0 = -0.05(13)(25) \text{ fm}$, respectively, where the first error is the statistical error and the second is the systematic error arising from the ERE expansion. We did not estimate the discretization error since we only have one lattice spacing. 

In this study, the effects of the coupled channels $\Xi_{cc}N$ and $\Sigma_c \Sigma_c$ are ignored in the scattering analysis. We computed the spectrum with both $\Lambda_c \Lambda_c$ and $\Xi_{cc}N$ operators, and observed that these two types of operators do not mix with each other. Therefore, we opted not to include the $\Xi_{cc}N$ channel in the scattering analysis. Furthermore, since the energy range explored is well below the $\Sigma_c \Sigma_c$ threshold, this channel is also ignored. To adequately investigate the coupled channel effects, more energy levels in the finite volume need to be calculated using more operators that interpolate the three channels with various momentum combinations. In addition, the $\Xi_{cc}N \pi$ three particle scattering needs to be included since its threshold is below the $\Sigma_c\Sigma_c$ threshold in our ensembles. These tasks pose significant numerical and theoretical challenges, necessitating further efforts in the future to address these issues.
\begin{center}
    \textbf{Acknowledgments}
\end{center}

We thank the CLQCD collaborations for providing us the gauge configurations~\cite{CLQCD:2023sdb}, which are generated on the HPC Cluster of ITP-CAS, the Southern Nuclear Science Computing Center(SNSC), the Siyuan-1 cluster supported by the Center for High Performance Computing at Shanghai Jiao Tong University and the Dongjiang Yuan Intelligent Computing Center. Software \verb|Chroma|~\cite{Edwards:2004sx} and \verb|QUDA|~\cite{Clark:2009wm, Babich:2011np, Clark:2016rdz} are used to generate the configurations and solve the perambulators. We are grateful to Hongxin Dong, Feng-Kun Guo, Wei Kou, Xiaopeng Wang and Haobo Yan for valuable discussions. This work is supported by NSFC under Grant No. 12293060, 12293061, 12293063, 12175279, 12435002, 12322503 and 12175239. L.~Liu also acknowledges the Strategic Priority Research Program of the Chinese Academy of Sciences with Grant No. XDB34030301 and Guangdong Major Project of Basic and Applied Basic Research No. 2020B0301030008. We also acknowledge the science and education integration young faculty project of University of Chinese Academy of Sciences YSBR-101. 

\bibliography{ref}

\end{document}